# Spin-orbit Torque Switching in an All-Van der Waals Heterostructure

Inseob Shin[1,*], Won Joon Cho[2,*], Eun-Su An[1,3], Sungyu Park[3], Hyeon-Woo Jeong[1], Seong Jang[1], Woon Joong Baek[2], Seong Yong Park[2], Dong-Hwan Yang[4], Jun Ho Seo[1], Gi-Yeop Kim[4], Mazhar N. Ali[5], Si-Young Choi[4], Hyun-Woo Lee[1,6], Jun Sung Kim[1,3], Sungdug Kim[2] and Gil-Ho Lee[1,6,†]

[1] Department of Physics, Pohang University of Science and Technology, Pohang 37673, Republic of Korea

[2] Material Research Center, Samsung Advanced Institute of Technology (SAIT), Samsung Electronics Co., Ltd, 130 Samsung-ro, Yeongtong-gu, Suwon-si, Gyeonggi-do 16678, Republic of Korea

[3] Center for Artificial Low Dimensional Electronic Systems, Institute for Basic Science (IBS), Pohang 37673, Korea

[4] Department of Materials Science and Engineering, Pohang University of Science and Technology, 77 Cheongam-Ro, Pohang 37673, Republic of Korea

[5] Max Plank Institute for Microstructure Physics, Weinberg 2, Halle (Saale) 06120, Germany

[6] Asia Pacific Center for Theoretical Physics, 77 Cheongam-Ro, Pohang 37673, Republic of Korea

* Equal contribution

† Correspondence and requests for materials should be addressed to G.-H.L. (lghman@postech.ac.kr).

**Current-induced control of magnetization in ferromagnets using spin-orbit torque (SOT) has drawn attention as a new mechanism for fast and energy efficient magnetic memory devices. Energy-efficient spintronic devices require a spin-current source with a large SOT efficiency ($\xi$) and electrical conductivity ($\sigma$), and an efficient spin injection across a transparent interface. Herein, we use single crystals of the van der Waals (vdW) topological semimetal WTe$_2$ and vdW ferromagnet Fe$_3$GeTe$_2$ to satisfy the requirements in their all-vdW-heterostructure with an atomically sharp interface. The results exhibit values of $\xi \approx 4.6$ and $\sigma \approx 2.25 \times 10^5$ $\Omega^{-1}m^{-1}$ for WTe$_2$. Moreover, we obtain the significantly reduced switching current density of $3.90 \times 10^6$ A/cm$^2$ at 150 K, which is an order of magnitude smaller than those of conventional heavy-metal/ferromagnet thin films. These findings highlight that engineering vdW-type topological materials and magnets offers a promising route to energy-efficient magnetization control in SOT-based spintronics.**

Spintronics, a next-generation information technology, is based on effective spin-current generation and injection. Spin transfer torque (STT) induced by spin-polarized charge current injection across one ferromagnet (FM) layer into another has been successfully employed for the effective manipulation of magnetization, leading to the recent commercial STT-based magnetic memory solutions [1]. Spin-orbit torque (SOT), which uses out-of-plane spin current generated from in-plane charge current in high spin Hall effect (SHE) materials, can realize a more energy-efficient manipulation of magnetization and is reaching commercial maturity [2-4]. Thus far, various high spin orbit coupling (SOC) materials, including heavy metals, topological insulators (TIs)[5-7], and recently, topological semimetals (TSMs)[8-10], have been studied to maximize their spin Hall angle, $\theta_{\text{SH}} = |J_s|/|J_c|$, a measure of their efficiency at converting charge current density $J_c$ to spin current density $J_s$. Also, the interface engineering between the layers of high-SHE and FM materials has been investigated to maximize the spin transparency, $T_{\text{int}}$, across the interface [11-18]. The key challenge for efficient SOT spintronic devices is to maximize the SOT efficiency, $\xi = \theta_{\text{SH}} \cdot T_{\text{int}}$ [19].

Recent rapid developments in van der Waals (vdW) materials and their heterostructures provide new opportunities for improved spintronic functionalities. A large group of topological materials was identified in vdW structures with an atomically flat surface. Also recently, FM vdW materials have been discovered [20, 21], some of which exhibit a relatively high Curie temperature ($T_C$) and perpendicular magnetic anisotropy (PMA) [22], which is important for FM layers in spintronic devices. Thus far, topological and FM vdW materials have been used as one of the constituent layers, either for spin-current generation or as the FM layer, *e.g.*, a vdW TI $Bi_2Se_3$ with a deposited FM CoFeB layer [7] or a vdW FM $Fe_3GeTe_2$ (FGT) with a deposited heavy-metal Pt layer [23, 24]. While current-induced magnetization control was successfully demonstrated in both cases, SOT performance in all-vdW heterostructures has yet to be explored. Here, using a vdW heterostructures of the topological semimetal $WTe_2$ and ferromagnet $Fe_3GeTe_2$, we show efficient current-induced magnetization switching with a much smaller switching current and power dissipation, as compared to conventional SOT devices. These observations highlight that all-vdW heterostructure with vdW TSMs and ferromagnets provide a promising architecture for SOT-based spintronic devices.

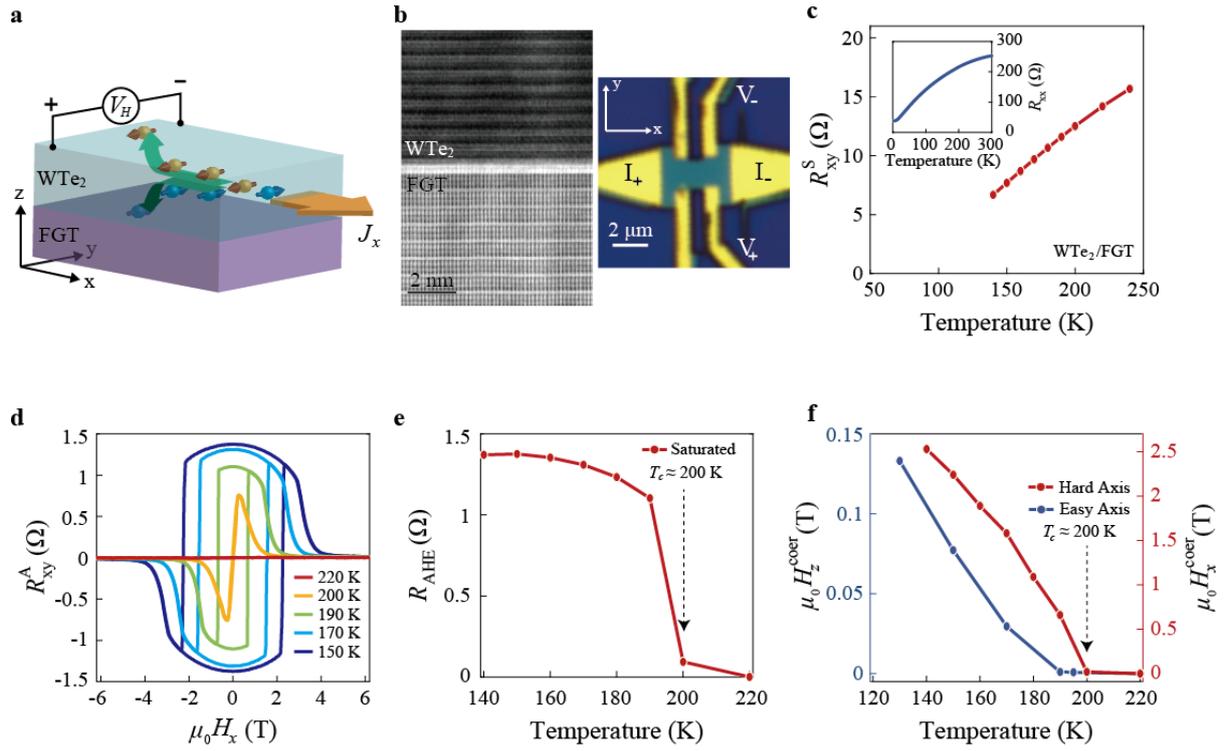

**Fig. 1 | Structure of WTe$_2$/Fe$_3$GeTe$_2$ (FGT) device and electric and magnetic properties of FGT. a**, Device schematics of an WTe$_2$/FGT heterostructure. The charge current ($J_x$, yellow arrow) applied to WTe$_2$ generates a spin current that is injected into the FGT along the -$z$-direction. By conservation of angular momentum, this spin current exerts a torque on the magnetization ($M$) of the FGT. Ball symbols with arrows represent electrons with their spin, respectively. **b**, Cross-sectional annular bright field-scanning transmission electron microscopy image of WTe$_2$/FGT heterostructure viewed from [120] directions of FGT (left panel) and optical image of the Hall bar device (right panel). The current is applied along the $x$-axis, and the Hall voltage is measured along the $y$-axis. **c**, Temperature dependence of symmetric part of Hall resistance at zero field $R_{xy}^S$ and longitudinal resistance $R_{xx}$ (inset) of the WTe$_2$/FGT Hall bar device. **d**, Anti-symmetrized Hall resistance $R_{xy}^A$ as a function of an in-plane magnetic field along $x$-axis $H_x$ at different temperatures. **e,f**, Temperature dependence of anomalous Hall

resistance $R_{AHE}$ (**e**), easy-axis ($H_z^{coer}$) and hard-axis ($H_x^{coer}$) coercive field (**f**) of the device. The critical temperature $T_c$ is approximately 200 K.

**Basic characteristics of the device**

Figure 1a schematically shows the spin Hall effect in WTe$_2$ producing a pure spin current, which is injected into the FGT layer, and exerts an SOT on the magnetization of FGT. In this study, we fabricated an all-vdW heterostructure consisting of WTe$_2$ (12.6 nm)/FGT (7.3 nm) using a dry transfer technique (see Fig. S1). The exfoliation of WTe$_2$ and FGT crystals, and transfer processes were performed in an inert Argon atmosphere glovebox. After capping the WTe$_2$/FGT stack with a 2.6-nm-thick aluminium oxide layer without exposing it to air, we patterned the stack into a Hall bar shape; this was followed by electrode deposition, as shown in Fig. 1b (see Methods and Fig. S2). The cross-sectional scanning transmission electron microscopic image of a representative WTe$_2$/FGT stack clearly shows the atomic layers of WTe$_2$ and FGT with the atomically sharp interface between them. This confirms that vdW stacking produces a clean interface without any residue or intermixing. Note that the brighter interfacial contrast is due to the atomic misalignment between [010] orientation of WTe$_2$ and [120] orientation of FGT by ~5°.

We first discuss the temperature and magnetic characteristics of the device. Figure 1c shows the temperature dependence of the symmetrized Hall resistance ($R_{xy}^S$) and longitudinal resistance ($R_{xx}$) of the device, the former of which will be used later to estimate electronic temperature at a high bias current (see the temperature dependence of the WTe$_2$ layer in Fig. S3). Figures 1d–f present the magnetic properties of the device. The anti-symmetrized Hall resistance $R_{xy}^A(H_x) = [R_{xy}(H_x) - R_{xy}(-H_x)]/2$ exhibits hysteretic behaviour with the in-plane magnetic field $H_x$, owing to a slight misalignment of the magnetic field. This is a typical hard-axis hysteresis loop for the materials with perpendicular magnetic anisotropy (PMA). As the temperature increases, the hysteresis gradually disappears, and the anomalous Hall resistance $R_{AHE}$, *i.e.*, $R_{xy}^A$ at $H_x = 0$, decreases and eventually vanishes at the critical temperature $T_c \sim 200$ K (Fig. 1e), which is close to the bulk $T_c = 205\sim220$ K [22, 25]. Hard- and easy-axis coercive fields ($H_x^{coer}$ and $H_z^{coer}$, respectively) also vanish at a similar $T_c$, as shown in Fig. 1f. The $T_c$ of the device is comparable to those measured in FGT of similar thickness in previous studies [22], thereby indicating no significant degradation of FGT during the device fabrication. Here, $H_z^{coer}$ is measured in another FGT device (7.86 nm) of the same $T_c \sim 200$ K (see Fig. S4).

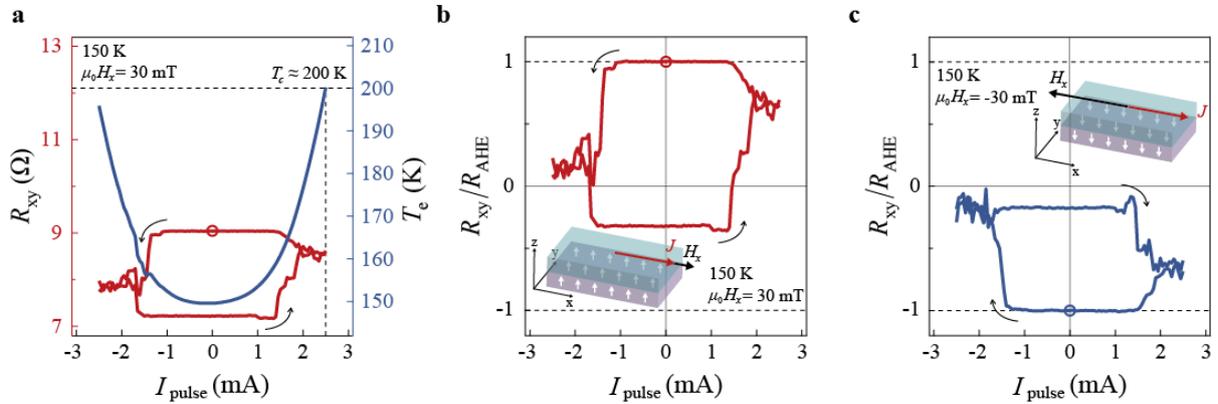

**Fig. 2 | Current-induced magnetization switching in WTe$_2$/Fe$_3$GeTe$_2$ (FGT) heterostructure device. a**, Hall resistance $R_{xy}$ (red line) measured after applying 10-ms-long current pulses of height $I_{pulse}$ in the presence of an in-plane magnetic field $\mu_0 H_x = 30$ mT parallel to the charge current at 150 K. $\mu_0$ is a vacuum permeability. The initial state of FGT is saturated

by the up-state ($M \parallel +z$-axis), which is denoted by a dot. Arrows indicate the $I_{pulse}$-sweep direction, and the switching polarity is anticlockwise. The electron temperature $T_e$ (blue line) during the current pulse is plotted as a function of $I_{pulse}$. The dashed lines represent the critical temperature $T_c \sim 200$ K of FGT and the corresponding $I_{pulse}$. **b,c**, The ratio of $R_{xy}$ to the anomalous Hall resistance $R_{AHE}$ measured with a sweeping $I_{pulse}$ under $\mu_0 H_x$ of 30 mT (**b**) and -30 mT (**c**) at 150 K. The switching polarity is anticlockwise for **b** and clockwise for **c**. Dots denote the initial magnetization states of FGT. The horizontal dashed lines represent the states of saturated magnetization of FGT.

**Results**

**Current-induced magnetization switching.** We show that the SOT that originated from the charge current flowing through $WTe_2$ can deterministically switch the perpendicular FGT magnetization of the device in the presence of $H_x$. To measure current-induced magnetization switching, we first applied a 10-ms-long current pulse of height $I_{pulse}$, and we subsequently measured $R_{xy}$ with a small bias current of 10 μA. As shown in Fig. 2a, the switching of $R_{xy}$ occurs at $I_{pulse} \sim \pm 1.5$ mA with an anticlockwise switching polarity for a positive $H_x$, which indicates a positive Hall angle in $WTe_2$ [26, 27]. Multiple sweeps of $I_{pulse}$ show consistent and stable switching loops, which will be discussed later (see Fig. S5). Unlike conventional FMs (e.g., Co or CoFeB), which have a $T_c$ that is considerably higher than the operation temperature, FGT has a relatively low $T_c$, such that the Joule heating effect may significantly alter the device performance. Thus, we have further analysed the elevation of the electronic temperature ($T_e$), owing to Joule heating from the current pulses (blue curve in Fig. 2a). A slight misalignment between the Hall electrodes in the device introduces a small $R_{xx}$ component in the measured voltage. By comparing it with the $R_{xx}(T)$ curve, which was measured beforehand using a small bias current of 10 µA (Fig. 1c), we were able to monitor the electronic temperature $T_e$ during the current pulses. While the rise of $T_e$ ($\Delta T_e$ ~10 K) is not significant at the switching $I_{pulse}$ (~±1.5 mA), $T_e$ could approach $T_c$ for the maximum $I_{pulse}$ of 2.5 mA. For a stable measurement, we limited the maximum $I_{pulse}$, such that $T_e$ did not exceed $T_c$. Figures 2b–c show the switching of the normalized Hall resistance ($R_{xy}/R_{AHE}$) with opposite directions of $H_x$: $\mu_0 H_x$=+30 mT in Fig. 2b and $\mu_0 H_x$=-30 mT in Fig. 2c. As expected for the current-induced magnetization switching of PMA magnets by SOT, the switching polarity is opposite when reversing the direction of $H_x$ [2]. The observation that $R_{xy}/R_{AHE}$ did not reach -1 for $\mu_0 H_x$ = +30 mT or +1 for $\mu_0 H_x$ = -30 mT can be explained by current spreading in the Hall bar device [5] and/or multi-domain formation in FGT, owing to the Joule heating of $T_e$ close to $T_c$, which will be discussed in further detail below [23].

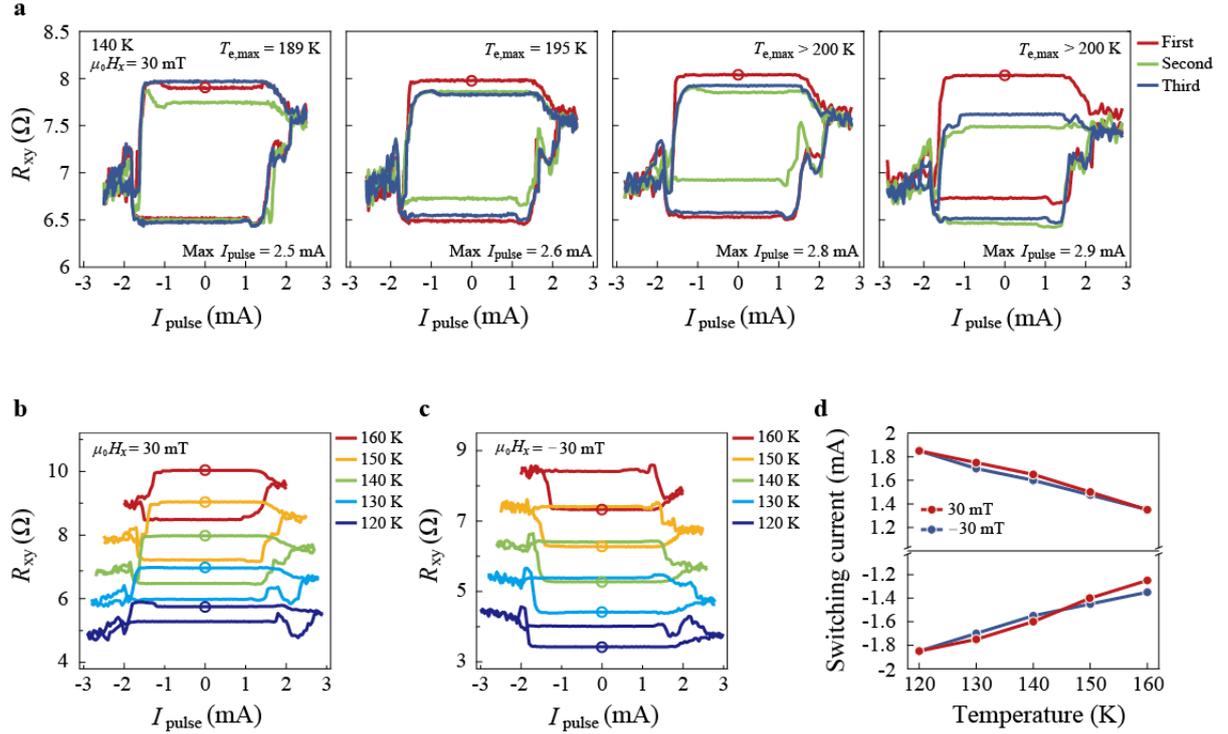

**Fig. 3 | Current pulse height $I_{pulse}$ and temperature dependence of current-induced magnetization switching. a**, Hall resistance $R_{xy}$ under three consecutive sweeps of $I_{pulse}$ with in-plane magnetic field $\mu_0 H_x$ = 30 mT at $T$ = 140 K. The maximum value of $I_{pulse}$ for each case is varied from 2.5 to 2.9 mA. The initial state of $Fe_3GeTe_2$ (FGT) is denoted by circles. For each case, the maximum electronic temperature $T_{e,max}$ was estimated as the electron temperature of the device during current pulses of maximum $I_{pulse}$. **b,c**, $R_{xy}$ under sweeps of $I_{pulse}$ at different temperatures with $\mu_0 H_x$ = 30 mT (**b**) and $\mu_0 H_x$ = -30 mT (**c**). The initial state of FGT is denoted by dot symbols. **d**, Switching current as a function of temperature for $\mu_0 H_x$ = 30 mT (red) and $\mu_0 H_x$ = -30 mT (blue).

**Joule heating effect.** To further investigate the effect of Joule heating on the stability of the switching behaviour, we measured current-induced switching under three consecutive sweeps of current pulses with various maximum $I_{\text{pulse}}$ (Fig. 3a). When the maximum $I_{\text{pulse}}$ is 2.5 mA, the maximum electronic temperature $T_{\text{e,max}} = 189$ K does not exceed $T_{\text{c}} = 200$ K for FGT, and three consecutive sweeps show a relatively consistent current-induced switching. However, as the maximum $I_{\text{pulse}}$ increases to 2.9 mA, $T_{\text{e,max}}$ becomes comparable to, or exceeds, $T_{\text{c}}$, and the current-induced switching behaviour becomes more inconsistent for consecutive sweeps. For these cases, the magnetization of the device cannot return to its original state. This occurs because significant thermal fluctuations induced by a large bias current randomise the magnetic domains of FGT. This also emphasises that the Joule heating issue is important to consider when using vdW FMs with a low $T_{\text{c}}$ [28], such as $Cr_2Ge_2Te_6$ or $CrI_3$ [20, 21].

**Temperature dependence of switching current.** The temperature dependence of current-induced switching is shown in Figs. 3b–c. The difference in the $R_{\text{xy}}$ value at opposite magnetizations decreases with a decrease in $T$, which is attributed to changes in the current distribution between $WTe_2$ and FGT. At lower $T$, the resistivity of $WTe_2$ decreases significantly, while that of FGT remains almost constant; therefore, less current flowing through FGT results in a smaller Hall voltage. Figure 3d summarizes the temperature dependence of the switching current (see Fig. S6–10). The reduction of the switching current with increasing $T$ is due to the simultaneous decrease in the magnetization of FGT and the easy-axis coercive field.

**Analysis of spin Hall angle and spin conductivity.** Generally, the charge-to-spin conversion efficiency is quantified by the SOT efficiency $\xi = \frac{4e}{\pi \hbar} M_{\text{FGT}} t_{\text{FGT}} \frac{H_z^{\text{coer}}}{J_{\text{SW}}}$,[24, 29-34] where $e$ is the electron charge, $\hbar$ is the reduced Planck constant, $M_{\text{FGT}}$ is the saturation magnetization of FGT, $t_{\text{FGT}}$ is the thickness of FGT, $H_z^{\text{coer}}$ is the easy-axis coercive field (Fig. 1f), and $J_{\text{SW}}$ is the switching current density of the device. From the parallel resistor model for $WTe_2$/FGT, we estimated that 60% of the total current flows through $WTe_2$ at $T_{\text{e}} \sim 160$ K ($T = 150$ K), which results in $J_{SW} = 3.90 \times 10^6$ A/cm². If taking $M_{FGT}$ = 240 emu/cm³ and $H_z^{coer}$ = 522 G at $T_e$~160 K, we obtain $\xi$ = 4.6. Here, $M_{FGT}$ is measured in a bulk crystal of FGT with $T_c$~210 K, which is similar to that of the SOT device discussed thus far (see Fig. S11).

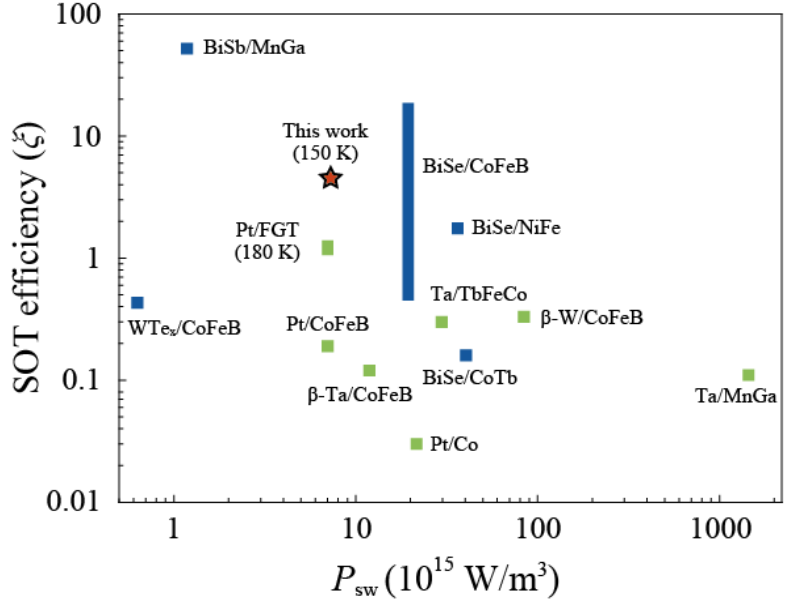

**Fig. 4 | Comparison of spin-orbit torque (SOT) efficiency and dissipation power density for magnetization switching.** Spin-orbit torque efficiency $\xi$ estimated from the current-induced magnetization switching and dissipation power density of the whole device at magnetization switching ($P_{sw}$) for devices based on heavy metals (green symbols) [2, 3, 24, 35-39] and topological materials (blue symbols) [5-7, 40, 41] at room temperature unless noted otherwise. The vertical length of the bar represents the range of the parameter. 'WTe$_x$' represents sputtered disordered WTe$_2$.

**Device performance comparison.** We now discuss the SOT performance of our WTe$_2$/FGT

device in comparison with previous SOT devices. First, we compare our WTe$_2$/FGT device to Pt/FGT devices that use the same FM material, FGT, but sputtered spin-current generation material, Pt. The switching current density $J_{sw}$ of our WTe$_2$/FGT ($3.90 \times 10^6$ A/cm$^2$) is an order of magnitude smaller than that of Pt/FGT devices. At a similar $T/T_c$~0.8, the $J_{sw}$ of Pt/FGT devices is $2.5 \times 10^7$ A/cm$^2$ at 180 K [24] and $1.2 \times 10^7$ A/cm$^2$ at 120 K [23]. Second, we compare our WTe$_2$/FGT device to WTe$_2$/Py that uses the same spin-current generation material, WTe$_2$, but with sputtered FM, Py. WTe$_2$/Py device had a SOT efficiency $\xi$=0.09–0.51 estimated with spin-torque ferromagnetic resonance measurement [27], which is an order of magnitude smaller than our result ($\xi$=4.6). These comparisons suggest that using WTe$_2$ as a spin-current generation material combined with FM layers through vdW stacking enhanced SOT performance, which can be attributed due to better spin transparency across the atomically sharp vdW interface. Lastly, WTe$_2$ is a TSM with a large conductivity ($2.5 \times 10^5$ $\Omega^{-1}$m$^{-1}$), which is an order of magnitude larger than that of Bi$_x$Se$_{1-x}$ TIs ($1.4- 9.4 \times 10^4$ $\Omega^{-1}$m$^{-1}$) [5, 7, 40, 42] or similar to that of narrow-gap TI Bi$_{0.9}$Sb$_{0.1}$ ($2.5 \times 10^5$ $\Omega^{-1}$m$^{-1}$) [6]. In addition, the conductivity of FGT ($2.69 \times 10^5$ $\Omega^{-1}$m$^{-1}$) is approximately 2–20 times smaller than that of conventional FMs, such as Co, Py, CoTb, and CoFeB ($0.5- 4.2 \times 10^6$ $\Omega^{-1}$m$^{-1}$) [5, 7, 35, 42]. Therefore, a larger portion of current flows through the spin-current generation layer, and the dissipation power density of the whole device at magnetization switching ($P_{sw}$) is lowered as summarised in Fig. 4, hence exhibiting improved energy-efficiency.

**Discussion**

Here we outline a few directions to pursue to further enhance the functionalities of all-vdW heterostructures. Currently, various two-dimensional ferromagnetic materials are actively being discovered and studied. Additional to $Fe_3GeTe_2$, which was used in this study, a series of $Fe_nGeTe_2$ ($3 \leq n \leq 5$) have been investigated theoretically and experimentally to enhance $T_c$ and the saturation magnetization [25]. In addition to its vdW layered and single crystalline structure, the topological nodal-line structure of spin-polarized bands of FGT may also provide better performance for spintronic applications with a large spin Hall angle [43]. Also, a theoretical study [44] and a related experiment [45] suggest that owing to the lowered geometrical symmetries in the single FGT layer, the current flow in FGT layer may induce SOT, which can contribute to magnetization switching. Moreover, $WTe_2$ has exhibited out-of-plane anti-damping torque, which was induced by a current bias along the low-symmetry axis [8, 9, 46]. This is owing to its broken screw-axis and glide plane symmetries at the surfaces, which is not allowed for conventional HMs. Such unconventional SOT can enable current-induced magnetization switching without an external magnetic field, which is highly desirable for practical applications and also demonstrated recently [41, 47]. Further optimization with novel topological materials, together with interface engineering, can lead to highly energy-efficient all-vdW spintronics devices.

**Methods**

A $WTe_2$ (FGT) single crystal was mechanically exfoliated to provide $WTe_2$ (FGT) thin flakes with a thickness of 10 ~ 15 nm (< 10 nm) on Gel-Film PF-30/17-X4 from Delphon Industries (silicon oxide wafer). The $WTe_2$ flakes were transferred onto the FGT flake to make $WTe_2$/FGT vdW heterostructures. All the exfoliation and transfer processes were performed in an inert Argon atmosphere glovebox to minimize the degradation of the materials and interfaces. To protect a stack of $WTe_2$/FGT during the fabrication process, aluminium oxide with a thickness of 2.6 nm was deposited via electron beam (e-beam) evaporation, without exposing the WTe$_2$/FGT stack to the air. The thickness of the WTe$_2$ and FGT flakes were confirmed using atomic force microscopy before the fabrication process. Electrodes were patterned using e-beam lithography, and in-situ Argon ion milling was used to eliminate the aluminium oxide of the WTe$_2$/FGT stack, followed by Cr (5 nm)/Au (35 nm) electrode deposition. Last, the WTe$_2$/FGT stack was shaped into Hall bar geometry using in-situ Argon ion milling, followed by aluminium oxide (50 nm) deposition to protect etched sides of the WTe$_2$/FGT stack from air exposure. Current direction of the WTe$_2$/FGT device is 23° off from *a*-axis of WTe$_2$ crystal.

**Acknowledgements**

This work was supported by Samsung Advanced Institute of Technology (SAIT). M.N.A. acknowledges support from the Alexander von Humboldt Foundation Sofia Kovalevskaja Award, the German Federal Ministry of Education and Research's MINERVA ARCHES Award, and the Max Planck Society. S.-Y.C. acknowledges the support of the Global Frontier Hybrid Interface Materials of the National Research Foundation of Korea (NRF) funded by the Ministry of Science and ICT (2013M3A6B1078872), and Korea Basic Science Institute (National research Facilities and Equipment Center) grant (2020R1A6C1020259) funded by the Ministry of Education. H.-W.L. acknowledges support from National Research Foundation of Korea (NRF) grant funded by the Korea government (MSIT) (No. 2020R1A2C2013484). J.S.K. was supported by the Institute for Basic Science (IBS) through the Center for Artificial Low Dimensional Electronic Systems (no. IBS-R014-D1) and by the National Research Foundation of Korea (NRF) through SRC (Grant No. 2018R1A5A6075964), the Max Planck-POSTECH Center for Complex Phase Materials (Grant No. 2016K1A4A4A01922028). G.-

H.L. was supported by National Research Foundation of Korea (NRF) funded by the Korean Government (Grant No. 2016R1A5A1008184, 2020R1C1C1013241, 2020M3H3A1100839), Samsung Science and Technology Foundation (project no. SSTF-BA1702-05), and Samsung Electronics Co., Ltd (IO201207-07801-01).

**Author contributions**

J.S.K., S.K. and G.-H.L. conceived and supervised the project. I.S., W.J.C., E.-S.A., S.J. and H.-W.J. fabricated the samples. I.S., E.-S.A., S.P. and S.J. performed transport experiments. M.N.A. grew $WTe_2$ crystals, and J.H.S. grew $Fe_3TeGe_2$ crystals and measured bulk magnetization. W.J.B., S.Y.P, D.-H.Y., G.-Y.K. and S.-Y.C. performed TEM measurements and analysis. I.S., W.J.C., H.-W.L., J.S.K. and G.-H.L. analysed transport data and wrote the paper with inputs from M.N.A., S.-Y.C. and S.K.

**References**

[1] A. Brataas, et al., Current-induced torques in magnetic materials, *Nat. Mater.*, **11,** 372-381, (2012)

[2] L. Liu, et al., Spin-torque switching with the giant spin Hall effect of tantalum, *Science*, **336,** 555-558, (2012)

[3] I. M. Miron, et al., Perpendicular switching of a single ferromagnetic layer induced by in-plane current injection, *Nature*, **476,** 189-193, (2011)

[4] J. Ryu, et al., Current-induced spin-orbit torques for spintronic applications, *Adv. Mater.*, **32,** e1907148, (2020)

[5] J. Han, et al., Room-temperature spin-orbit torque switching induced by a topological insulator, *Phys. Rev. Lett.*, **119,** 077702, (2017)

[6] N. H. D. Khang, et al., A conductive topological insulator with large spin Hall effect for ultralow power spin-orbit torque switching, *Nat. Mater.*, **17,** 808-813, (2018)

[7] M. Dc, et al., Room-temperature high spin-orbit torque due to quantum confinement in sputtered $Bi_xSe_{(1-x)}$ films, *Nat. Mater.*, **17,** 800-807, (2018)

[8] D. MacNeill, et al., Thickness dependence of spin-orbit torques generated by $WTe_2$, *Phys. Rev. B*, **96,** 054450, (2017)

[9] D. MacNeill, et al., Control of spin–orbit torques through crystal symmetry in $WTe_2$/ferromagnet bilayers, *Nat. Phys.*, **13,** 300-305, (2016)

[10] P. Li, et al., Spin-momentum locking and spin-orbit torques in magnetic nano-heterojunctions composed of Weyl semimetal $WTe_2$, *Nat. Commun.*, **9,** 3990, (2018)

[11] X. Fan, et al., Quantifying interface and bulk contributions to spin-orbit torque in magnetic bilayers, *Nat. Commun.*, **5,** 3042, (2014)

[12] X. Qiu, et al., Spin-orbit-torque engineering via oxygen manipulation, *Nat. Nanotech.*, **10,** 333-338, (2015)

[13] Y. W. Oh, et al., Field-free switching of perpendicular magnetization through spin-orbit torque in antiferromagnet/ferromagnet/oxide structures, *Nat. Nanotech.*, **11,** 878-884, (2016)

[14] J. C. Rojas-Sanchez, et al., Spin pumping and inverse spin Hall effect in platinum: the essential role of spin-memory loss at metallic interfaces, *Phys. Rev. Lett.*, **112,** 106602, (2014)

[15] W. Zhang, et al., Role of transparency of platinum–ferromagnet interfaces in determining the intrinsic magnitude of the spin Hall effect, *Nat. Phys.*, **11,** 496-502, (2015)

[16] W. Zhang, et al., Reduced spin-Hall effects from magnetic proximity, *Phys. Rev. B*, **91,** 115316, (2015)

[17] V. P. Amin and M. D. Stiles, Spin transport at interfaces with spin-orbit coupling: Phenomenology, *Phys. Rev. B*, **94,** 104420, (2016)

[18] S. C. Baek, et al., Spin currents and spin-orbit torques in ferromagnetic trilayers, *Nat. Mater.*, **17,** 509-513, (2018)

[19] F. Hellman, et al., Interface-induced phenomena in magnetism, *Rev. Mod. Phys.*, **89,** 025006, (2017)

[20] C. Gong, et al., Discovery of intrinsic ferromagnetism in two-dimensional van der Waals crystals, *Nature*, **546,** 265-269, (2017)

[21] B. Huang, et al., Layer-dependent ferromagnetism in a van der Waals crystal down to the monolayer limit, *Nature*, **546,** 270-273, (2017)

[22] Y. Deng, et al., Gate-tunable room-temperature ferromagnetism in two-dimensional $Fe_3GeTe_2$, *Nature*, **563,** 94-99, (2018)

[23] X. Wang, et al., Current-driven magnetization switching in a van der Waals ferromagnet $Fe_3GeTe_2$, *Sci. Adv.*, **5,** eaaw8904, (2019)

[24] M. Alghamdi, et al., Highly efficient spin-orbit torque and switching of layered ferromagnet $Fe_3GeTe_2$, *Nano. Lett.*, **19,** 4400-4405, (2019)

[25] J. Seo, et al., Nearly room temperature ferromagnetism in a magnetic metal-rich van der Waals metal, *Sci. Adv.*, **6,** eaay8912, (2020)

[26] B. Zhao, et al., Observation of charge to spin conversion in Weyl semimetal $WTe_2$ at room temperature, *Phys. Rev. Res.*, **2,** 013286, (2020)

[27] S. Shi, et al., All-electric magnetization switching and Dzyaloshinskii-Moriya interaction in $WTe_2$/ferromagnet heterostructures, *Nat. Nanotech.*, **14,** 945-949, (2019)

[28] Y. Shao, et al., The current modulation of anomalous Hall effect in van der Waals Fe$_3$GeTe$_2$/WTe$_2$ heterostructures, *Appl. Phys. Lett.*, **116,** 092401, (2020)

[29] T.-Y. Chen, et al., Tunable spin-orbit torque in Cu-Ta binary alloy heterostructures, *Phys. Rev. B*, **96,** 104434, (2017)

[30] P. P. Haazen, et al., Domain wall depinning governed by the spin Hall effect, *Nat. Mater.*, **12,** 299-303, (2013)

[31] M. Hayashi, et al., Quantitative characterization of the spin-orbit torque using harmonic Hall voltage measurements, *Phys. Rev. B*, **89,** 144425, (2014)

[32] J. Kim, et al., Layer thickness dependence of the current-induced effective field vector in Ta | CoFeB | MgO, *Nat. Mater.*, **12,** 240-245, (2013)

[33] A. Thiaville, et al., Dynamics of Dzyaloshinskii domain walls in ultrathin magnetic films, *Euro. Phys. Lett.*, **100,** 57002, (2012)

[34] L. Zhu, et al., Lack of simple correlation between switching current density and spin-orbit torque efficiency of perpendicularly magnetized spin-current generator/ferromagnet heterostructures, *arXiv:2101.10521*, (2021)

[35] L. Liu, et al., Current-induced switching of perpendicularly magnetized magnetic layers using spin torque from the spin Hall effect, *Phys. Rev. Lett.*, **109,** 096602, (2012)

[36] K. Meng, et al., Modulated switching current density and spin-orbit torques in MnGa/Ta films with inserting ferromagnetic layers, *Sci. Rep.*, **6,** 38375, (2016)

[37] H.-Y. Lee, et al., Enhanced spin–orbit torque via interface engineering in Pt/CoFeB/MgO heterostructures, *APL Mater.*, **7,** 031110, (2019)

[38] C.-F. Pai, et al., Spin transfer torque devices utilizing the giant spin Hall effect of tungsten, *Appl. Phys. Lett.*, **101,** 122404, (2012)

[39] Z. Zhao, et al., Spin Hall switching of the magnetization in Ta/TbFeCo structures with

bulk perpendicular anisotropy, *Appl. Phys. Lett.*, **106,** 132404, (2015)

[40] Y. Wang, et al., Room temperature magnetization switching in topological insulator-ferromagnet heterostructures by spin-orbit torques, *Nat. Commun.*, **8,** 1364, (2017)

[41] X. Li, et al., Large and robust charge-to-spin conversion in sputtered Weyl semimetal WTe$_x$ with structural disorder, *arXiv:2001.04054*, (2020)

[42] A. R. Mellnik, et al., Spin-transfer torque generated by a topological insulator, *Nature*, **511,** 449-451, (2014)

[43] K. Kim, et al., Large anomalous Hall current induced by topological nodal lines in a ferromagnetic van der Waals semimetal, *Nat. Mater.*, **17,** 794-799, (2018)

[44] O. Johansen, et al., Current control of magnetism in two-dimensional Fe$_3$GeTe$_2$, *Phys. Rev. Lett.*, **122,** 217203, (2019)

[45] K. Zhang, et al., Gigantic current control of coercive field and magnetic memory based on nanometer-thin ferromagnetic van der Waals Fe$_3$GeTe$_2$, *Adv. Mater.*, **33,** e2004110, (2020)

[46] B. Zhao, et al., Charge-spin conversion signal in WTe$_2$ van der Waals hybrid devices with a geometrical design, *Appl. Phys. Lett.*, **117,** 242401, (2020)

[47] I.-H. Kao, et al., Field-free deterministic switching of a perpendicularly polarized magnet using unconventional spin-orbit torques in WTe$_2$, *arXiv:2012.12388*, (2020)